\documentclass[jkps,twoside,twocolumn]{revtex4}
\usepackage{amsmath}
\usepackage{amssymb}

\begin{document}
\title[]{Supersymmetric Matrix Quantum Mechanics with Non-Singlet Sector}
\author{Nyong-Chol \surname{Kim}}
\email{nckim@khu.ac.kr} \affiliation{Department of Physics and
Research Institute for Basic Sciences, Kyung Hee University, Seoul,
130-701, Korea}
\author{Soonkeon \surname{Nam}}
\email{nam@khu.ac.kr} \affiliation{Department of Physics and
Research Institute for Basic Sciences, Kyung Hee University, Seoul,
130-701, Korea}
\author{Yunseok \surname{Seo}}
\email{yseo@ihanyang.ac.kr} \affiliation{Department of Physics and
Research Institute for Natural Sciences, Hanyang University, Seoul,
133-791, Korea}

\def\pp{\hbox{\ooalign{$\displaystyle\int$\cr$-$}}}
\def\ppa{\hbox{\ooalign{$-$\cr$\displaystyle\int_{-L/2}^{+L/2}$}}}

\begin{abstract}
We consider a supersymmetric matrix model which is related to the
non-critical superstring theory. We find new non-singlet terms in
the supersymmetric matrix quantum mechanics. The new non-singlet
terms give rise to nontrivial interactions. These new non-singlet
terms from fermions, can eliminate other non-singlet terms from
generators of $U(N)$ subalgebra and from time periodicity. The
non-singlet terms from the generators violate the T-duality on the
target space which is a circle. Therefore, we can retain the
T-duality with a process of the elimination.
\end{abstract}

\maketitle

\section{Introduction}
The matrix models have been used widely in many mathematical and
physical applications, such as combinatorics of graphs, topology,
integrable systems, string theory, theory of mesoscopic systems and
statistical mechanics on random surfaces \cite{GG, FGZ,
Polychronakos}. In this paper we will focus on the supersymmetric
matrix model which is related to the non-critical 2-dimensional
string theory. The worldsheet of the 2-dimensional string theory is
represented mathematically by random surfaces in matrix description
\cite{Klebanov, Das, Martinec}. The matrix description helps us to
understand the non-perturbative effects of the string theory.

The supersymmetric matrix quantum mechanics\cite{Ovrut} and the
bosonic matrix model with a time periodicity\cite{Boulatov} have
been studied previously. In this paper, we will focus on the
supersymmetric matrix quantum mechanics with non-singlet sector.
Such models are related to the 2-dimensional black hole\cite{KKK}.

In the matrix quantum mechanics, only the singlet sector has been
considered because the non-singlet terms are difficult to handle.
The non-singlet terms violate the T-duality on the target space.
These non-singlet terms are from generators of $U(N)$ subalgebra. In
this paper we construct new non-singlet terms from fermions. Our new
non-singlet terms prevent the old non-singlet terms from violating
the T-duality.

\section{A brief review}
In this section, firstly, we review the quantum mechanics of the
supersymmetric matrix model\cite{Ovrut}. Secondly, we review the
bosonic matrix model with periodic time condition\cite{Boulatov}.
Lastly, we review the non-singlet terms from generators of $U(N)$
subalgebra and review the violation of the T-duality on the target
space\cite{Gross0, Klebanov, Gross1, Gross2, Alexandrov, Koetsier}.

\subsection{Supersymmetric matrix quantum mechanics}
\subsubsection{Lagrangian and Hamiltonian}
We will use a time-dependent, $N\times N$, $d=1$, $\mathcal{N}=2$
Hermitian matrix superfield as follows :
%eq-1
\begin{equation}\Phi_{ij}(t)=M_{ij}(t)+i\theta_{1}\Psi_{1ij}(t)+i\theta_{2}\Psi_{2ij}(t)
+i\theta_{1}\theta_{2}F_{ij}(t),\end{equation} where $\theta_{1}$
and $\theta_{2}$ are real anticommuting parameters, $M_{ij}$ and
$F_{ij}$ are $N\times N$ bosonic Hermitian matrices and $\Psi_{1ij}$
and $\Psi_{2ij}$ are $N\times N$ fermionic Hermitian matrices. The
Lagrangian is
%eq-2
\begin{equation}L=\int d\theta_{1}d\theta_{2}\left\{\frac{1}{2}TrD_{1}\Phi
D_{2}\Phi+iW(\Phi)\right\},\end{equation} where the potential, $W$,
is a polynomial in $\Phi$,
%eq-3
\begin{equation}W(\Phi)=\sum_{n}b_{n}Tr\Phi^{n}.\end{equation}
In the above $b_{n}$ are real coupling parameters, and $D_{I}$ are
superspace derivatives,
%eq-4
\begin{equation}D_{I}=\frac{\partial}{\partial\theta_{I}}+i\theta_{I}\frac{\partial}{\partial
t}\,,\qquad (I=1, 2).\end{equation} In component fields, the
Lagrangian reads
%eq-5
\begin{equation}\begin{split}L=
&\sum_{ij}\left\{\frac{1}{2}(\dot{M}_{ij}\dot{M}_{ji}+F_{ij}F_{ji})+\frac{\partial
W(M)}{\partial M_{ij}}F_{ij}\right\} \\
&-\frac{i}{2}\sum_{ij}(\Psi_{1ij}\dot{\Psi}_{1ji}+\Psi_{2ij}\dot{\Psi}_{2ji})\\
&-i\sum_{ijkl}\Psi_{1ij}\frac{\partial^{2}W(M)}{\partial
M_{ij}\partial M_{kl}}\Psi_{2kl}.\end{split}\end{equation} The
equation of motion, $F_{ij}=-\frac{\partial W(M)}{\partial M_{ij}}$,
for the auxiliary matrix $F_{ij}$ makes the Lagrangian as follows :
%eq-6
\begin{equation}\begin{split}L=
&\sum_{ij}\left\{\frac{1}{2}\dot{M}_{ij}\dot{M}_{ji}-\frac{1}{2}\frac{\partial
W(M)}{\partial M_{ij}}\frac{\partial
W(M)}{\partial M_{ji}}\right\} \\
&-\frac{i}{2}\sum_{ij}(\Psi_{1ij}\dot{\Psi}_{1ji}+\Psi_{2ij}\dot{\Psi}_{2ji})\\
&-i\sum_{ijkl}\Psi_{1ij}\frac{\partial^{2}W(M)}{\partial
M_{ij}\partial M_{kl}}\Psi_{2kl}.\end{split}\end{equation} If we
turn off the fermionic terms in Eq.(6) then we get a pure bosonic
matrix model. The conjugate momenta to the matrices $M$, $\Psi_{1}$
and $\Psi_{2}$ are
%eq-7
\begin{equation}\begin{split}\Pi_{M_{ij}}=&\;\dot{M}_{ij},\\
\Pi_{\Psi_{1ij}}=&-\frac{i}{2}\Psi_{1ij},\\
\Pi_{\Psi_{2ij}}=&-\frac{i}{2}\Psi_{2ij}.\end{split}\end{equation}
The Legendre transformation for the Lagrangian with Eq.(7) gives us
following Hamiltonian,
%eq-8
\begin{equation}\begin{split}H=
&\sum^{N}_{ij}\left\{\frac{1}{2}\Pi_{M_{ij}}\Pi_{M_{ji}}+\frac{1}{2}\frac{\partial
W(M)}{\partial M_{ij}}\frac{\partial W(M)}{\partial
M_{ji}}\right\}\\
&+i\sum^{N}_{ijkl}\Psi_{1ij}\Psi_{2kl}\frac{\partial^{2}W(M)}{\partial
M_{ij}\partial M_{kl}}.\end{split}\end{equation} Let us introduce
the following relations for the canonical quantization of the
Hamiltonian,
%eq-9
\begin{equation}\begin{split}[\hat{\Pi}_{M_{ij}},
\hat{M}_{kl}]=&-i\delta_{ik}\delta_{jl},\\
\{\hat{\Psi}_{Iij},
\hat{\Psi}_{Jkl}\}=&\;\delta_{IJ}\delta_{ik}\delta_{jl},\end{split}\end{equation}
and the complex formation for the fermions,
%eq-10
\begin{equation}\begin{split}\hat{\Psi}=&\frac{1}{\sqrt{2}}(\hat{\Psi}_{1}+i\hat{\Psi}_{2})
,\\
\hat{\bar{\Psi}}=&\frac{1}{\sqrt{2}}(\hat{\Psi}_{1}-i\hat{\Psi}_{2}).\end{split}\end{equation}
The anti-commutator for $\hat{\bar{\Psi}}$ and $\hat{\Psi}$ is
$\{\hat{\bar{\Psi}}_{ij}, \hat{\Psi}_{kl}\}=\delta_{ik}\delta_{jl}.$
Therefore, the Hamiltonian has a final form,
%eq-11
\begin{equation}\begin{split}\hat{H}=
&\sum^{N}_{ij}\left\{\frac{1}{2}\hat{\Pi}_{M_{ij}}\hat{\Pi}_{M_{ji}}+\frac{1}{2}\frac{\partial
W(\hat{M})}{\partial\hat{M}_{ij}}\frac{\partial W(\hat{M})}{\partial
\hat{M}_{ji}}\right\}\\
&+\frac{1}{2}\sum^{N}_{ijkl}[\hat{\bar{\Psi}}_{ij},\hat{\Psi}_{kl}]\frac{\partial^{2}W(\hat{M})}{\partial
\hat{M}_{ij}\partial\hat{M}_{kl}}.\end{split}\end{equation}

\subsubsection{Unitary transformation and the singlet sector}
Now, let us take the unitary transformation
%eq-12
\begin{equation}\Phi(t)\rightarrow
U^{\dag}(t)\Phi(t)U(t)\end{equation} for the matrix superfield
$\Phi_{ij}(t)$ in Eq.(1). This unitary transformation makes the
bosonic matrix $M$ which is diagonal as
%eq-13
\begin{equation}M_{ij}=\sum^{N}_{k}U^{\dag}_{ik}\lambda_{k}U_{kj}.\end{equation}
However, in general, the fermionic matrix $\Psi$ is not be
diagonalized simultaneously. Although the off-diagonal result of
$\chi_{kl}$ for $k\neq l$, the only diagonal elements
$\chi_{kk}\equiv\chi_{k}$ for $k=l$ have been mostly used in the
supersymmetric matrix quantum mechanics up to date. It has diagonal
formation such as
%eq-14
\begin{equation}\Psi_{ij}=\sum^{N}_{kl}U^{\dag}_{ik}\chi_{k}U_{kj},\end{equation}
by the unitary transformation.

In general case, the supersymmetric matrix quantum mechanics has
so-called non-singlet terms related to the off-diagonal elements in
the fermionic matrices. This makes us define a ``rotated" fermion
matrix
%eq-15
\begin{equation}\chi=U\Psi U^{\dag}.\end{equation} We emphasize again that the unitary operator $U$ diagonalizes the $M$, but the $\chi$ is not diagonalized
simultaneously in general. However, the states $|s\rangle$ on the
$U(N)$-singlet sector in the Hilbert space, are annihilated by
$\frac{\partial}{\partial U_{ij}}$ and are also annihilated by
$\hat{\chi}_{ij}$ where $i\neq j$. Thus, if we concentrate those
states to the singlet sector then we are able to take only diagonal
terms of the fermionic matrices. In this case, the superfield,
Eq.(1), is transformed into as follows :
%eq-16
\begin{equation}\bigl(U^{\dag}\Phi(t)U\bigr)_{ii}\equiv\lambda_{i}(t)+i\theta_{1}\chi_{1i}(t)+i\theta_{2}\chi_{2i}(t)
+i\theta_{1}\theta_{2}f_{i}(t),\end{equation} where
$f_{i}\equiv(U^{\dag}FU)_{ii}$. Another example for Eq.(16) is in
Ref.\cite{Tonder}.

\subsubsection{Effective Lagrangian}
Now, we can construct a Hamiltonian for the singlet sector,
$\hat{H}_{s}$, such that $\hat{H}|s\rangle=\hat{H}_{s}|s\rangle$.
The ordinary Hamiltonian, $\hat{H}$, is given by Eq.(11) and the
singlet-Hamiltonian is
%eq-17
\begin{equation}\begin{split}\hat{H}_{s}=&\sum_{i}\left\{\frac{1}{2}\hat{\Pi}^{2}_{\lambda_{i}}+i\frac{\partial
w}{\partial\lambda_{i}}\hat{\Pi}_{\lambda_{i}}+\frac{1}{2}\left(\frac{\partial
W}{\partial\lambda_{i}}\right)^{2}+\frac{1}{2}\frac{\partial
w}{\partial\lambda_{i}}\frac{\partial
W}{\partial\lambda_{i}}\right\}\\
&+\frac{1}{2}\sum_{ij}[\bar{\chi}_{i},\chi_{j}]\frac{\partial^{2}W}{\partial\lambda^{2}_{i}},\end{split}\end{equation}
where $\chi_{i}\equiv\chi_{ii}$ and
%eq-18
\begin{equation}w=-\sum_{i}\sum_{j\neq
i}\ln|\lambda_{i}-\lambda_{j}|.\end{equation}

The effective Lagrangian, $L_{s}$, for $\hat{H}_{s}$ is given by the
Legendre transformation and the gaussian integration for the
$[d\Pi_{\lambda}]$ in following partition function \cite{Ovrut},
%eq-19
\begin{equation}\begin{split}Z_{N}(b_{n})=&\int[d\Pi_{\lambda}][d\lambda][d\chi][d\bar{\chi}]\exp\left[i\int
dt\sum_{ij}\bar{\chi_{i}}\chi_{j}\frac{\partial^{2}w}{\partial\lambda_{i}\partial\lambda_{j}}\right]\\
&\times\exp\left[i\int dt\sum_{i}\left\{\Pi_{\lambda_{i}}\dot{\lambda}_{i}-i\bar{\chi}_{i}\dot{\chi}_{i}-H_{s}\right\}\right]\\
=&\int[d\lambda][d\chi][d\bar{\chi}]\exp\left(i\int dt
L_{s}\right).\end{split}\end{equation} Thus the effective Lagrangian
for the singlet sector in the supersymmetric matrix case is
%eq-20
\begin{equation}\begin{split}L_{s}=&\sum_{i}\left\{\frac{1}{2}\dot{\lambda}^{2}_{i}-\frac{1}{2}\left(\frac{\partial W_{eff}}{\partial\lambda_{i}}\right)^{2}-\frac{i}{2}(\bar{\chi}_{i}\dot{\chi}_{i}-\dot{\bar{\chi}}_{i}\chi_{i})\right\}\\
&-\sum_{ij}\bar{\chi}_{i}\chi_{j}\frac{\partial^{2}W_{eff}}{\partial\lambda_{i}\partial\lambda_{j}},\end{split}\end{equation}
where the effective potential is
%eq-21
\begin{equation}W_{eff}(\lambda_{i})=W(\lambda_{i})+w(\lambda_{i}).\end{equation}
After inserting Eq.(21) into Eq.(20) we can rearrange terms to
obtain the final effective Lagrangian,
%eq-22
\begin{equation}\begin{split}L_{s}=&\sum_{i}\left\{\frac{1}{2}\dot{\lambda}^{2}_{i}-\frac{1}{2}\left(\frac{\partial W}{\partial\lambda_{i}}\right)^{2}-\frac{\partial w}{\partial\lambda_{i}}\frac{\partial W}{\partial\lambda_{i}}\right.\\
&\left.-\frac{1}{2}\left(\frac{\partial w}{\partial\lambda_{i}}\right)^{2}-\frac{i}{2}(\bar{\chi}_{i}\dot{\chi}_{i}-\dot{\bar{\chi}}_{i}\chi_{i})\right\}\\
&-\sum_{ij}\left\{\frac{\partial^{2}W}{\partial\lambda_{i}\partial\lambda_{j}}\bar{\chi}_{i}\chi_{j}+\frac{\partial^{2}w}{\partial\lambda_{i}\partial\lambda_{j}}\bar{\chi}_{i}\chi_{j}\right\}.\end{split}\end{equation}
A detailed derivation for the effective Lagrangian is given in
Appendix B of Ref.\cite{Ovrut}.

\subsection{The bosonic matrix quantum mechanics with time periodicity}
\subsubsection{Constraint from time periodicity}
We take a partition function for a time dependent and periodic
bosonic matrix,
%eq-23
\begin{equation}M_{ij}(t)=M_{ij}(t+\beta),\end{equation} as follows
:
%eq-24
\begin{equation}\begin{split}Z_{N}=&\int_{M(0)=M(\beta)}\mathcal{D}M\\
&\times\exp\left[-N\text{Tr}\int^{\beta}_{0}dt\left\{\frac{1}{2}\dot{M}^{2}+V(M)\right\}\right].\end{split}\end{equation}
The unitary transformation
$M_{ij}=\sum^{N}_{k}U^{\dag}_{ik}\lambda_{k}U_{kj},$ makes
$\text{Tr}\dot{M}^{2}$ in Eq.(24) as follows :
%eq-25
\begin{equation}\text{Tr}\dot{M}^{2}=\sum^{N}_{i}\dot{\lambda}^{2}_{i}+\sum_{i\neq
j}(\lambda_{i}-\lambda_{j})^{2}|A_{ij}(t)|^{2},\end{equation} where
%eq-26
\begin{equation}A_{ij}(t)=\bigl(\dot{U}(t)U^{\dag}(t)\bigr)_{ij}.\end{equation}
The measure $\mathcal{D}M$ is transformed into
%eq-27
\begin{equation}\mathcal{D}M_{ij}=[dA_{ij}]\prod_{i}[d\lambda_{i}]\prod_{i<j}(\lambda_{i}-\lambda_{j})^{2}.\end{equation}
In general, the measure of $[dA]$ can be dropped because the result
of the integral for $[dA]$ becomes trivial gauge volume factor.
However, in the case with the periodic time condition, we must keep
the measure of $[dA]$ because the integral for $[dA]$ gives us
non-trivial and important terms. Now, let us look into the
non-trivial result from the time periodicity.

The periodic time condition in Eq.(23) gets the diagonal elements
$\lambda_{i}$ of $M_{ij}(t)$ to be
%eq-28
\begin{equation}\lambda_{k}(t+\beta)=\sum^{N}_{j}\mathcal{P}_{kj}\lambda_{j}(t)\mathcal{P}^{-1}_{jk}\end{equation}
where $\mathcal{P}$ is an operator which makes the unitary operator,
$U(t)$,
%eq-29
\begin{equation}U(t+\beta)=\mathcal{P}U(t).\end{equation} The
connection $A_{ij}(t)$ is an independent variable in Eq.(27).
However, in the case of time periodicity, $A_{ij}(t)$ is not an
independent variable but is constrained the constraint with
$\mathcal{P}$ such that
%eq-30
\begin{equation}\widehat{T}\exp
i\int^{\beta}_{0}dtA(t)=\mathcal{P}^{-1},\end{equation} where
$\widehat{T}$ is a time-ordering operator. This constraint
contributes a delta-function to the measure Eq.(27) :
%eq-31
\begin{equation}\begin{split}\mathcal{D}M_{ij}(t)=&dA_{ij}(t)\prod_{i}d\lambda_{i}(t)\prod_{t\in[0,\beta]}\Delta^{2}(\lambda(t))\\
&\times\delta\bigl(\widehat{T}\exp
i\int^{\beta}_{0}dtA(t),\mathcal{P}^{-1}\bigr),\end{split}\end{equation}
where $\Delta(\lambda)=\prod_{k>m}(\lambda_{k}-\lambda_{m})$ is the
Vandermond determinant. The delta-function can relate to the
irreducible representation for the operator. The relation is
%eq-32
\begin{equation}\delta(U,\mathcal{P}^{-1})=\sum_{\mathcal{R}}d_{\mathcal{R}}\chi_{\mathcal{R}}(\mathcal{P}U),\end{equation}
where $d_{\mathcal{R}}$ is the dimension of the $\mathcal{R}$'th
representation and $\chi_{\mathcal{R}}$ is the character
%eq-33
\begin{equation}\begin{split}\chi_{\mathcal{R}}\bigl(\widehat{T}&\exp
\left(i\int^{\beta}_{0}A(t)dt\right)\bigr)\\
&=\text{Tr}_{\mathcal{R}}\left[\widehat{T}\exp
\left(i\int^{\beta}_{0}dt\sum_{i,j}A_{ij}\hat{\tau}^{\mathcal{R}}_{ij}\right)\right],\end{split}\end{equation}
where $\hat{\tau}^{\mathcal{R}}_{ij}$ is a generator of $U(N)$ in
the $\mathcal{R}$'th representation.

\subsubsection{Non-singlet terms from the constraint}
In the case of free boundary conditions (\textit{i.e.} when $M(0)$
is independent of $M(\beta)$) the singlet partition function
describes non-interacting fermions. After integration over all
angular variables, only two Vandermonde determinants at the ends of
the interval remain. They are $\Delta(\lambda(0))$,
$\Delta(\lambda(\beta))$ and these terms assure the anti-symmetry of
wave functions. In the case of the periodic boundary conditions, one
should be more careful and use delta functions to match the
eigenvalues as follows :
%eq-34
\begin{equation}f\bigl(\lambda(\beta)\bigr)=\int\prod_{k}d\lambda_{k}(0)\Delta^{2}\bigl(\lambda(0)\bigr)f\bigl(\lambda(0)\bigr)\delta\bigl(\lambda(0)-\lambda(\beta)\bigr).\end{equation}
Now, let us write the final partition function by the gaussian
integral for $dA_{ij}$ with preceding equations and relations. The
partition function is
%eq-35
\begin{equation}\begin{split}Z_{N}&=\frac{1}{N!}\sum_{\{\mathcal{P}\}}(-1)^{\mathcal{P}}\\
&\times\int\prod^{N}_{i}d\lambda_{i}\exp\left\{-N\sum_{i}\int^{\beta}_{0}dt\left(\frac{1}{2}\dot{\lambda}^{2}_{i}+V(\lambda_{i})\right)\right\} \\
&\times\sum_{\mathcal{R}}d_{\mathcal{R}}\text{Tr}_{\mathcal{R}}\left[\widehat{T}\exp\left\{\frac{1}{4N}\int^{\beta}_{0}dt\sum_{i\neq
j}\frac{\hat{\tau}^{\mathcal{R}}_{ij}\hat{\tau}^{\mathcal{R}}_{ji}}{(\lambda_{i}-\lambda_{j})^{2}}\right\}\mathcal{P}\right],\end{split}\end{equation}
where $\frac{1}{N!}\sum_{\{\mathcal{P}\}}(-1)^{\mathcal{P}}$ is the
standard anti-symmetrizator, the skew-symmetry of the Vandermond
determinant,
%eq-36
\begin{equation}\Delta\bigl(\mathcal{P}\lambda(0)\mathcal{P}^{-1}\bigr)=(-1)^{\mathcal{P}}\Delta\bigl(\lambda(0)\bigr).\end{equation}
Thus we have the Hamiltonian as follows :
%eq-37
\begin{equation}\begin{split}H_{\mathcal{R}}=P_{\mathcal{R}}\sum^{N}_{i}&\left\{-\frac{1}{2N}\frac{\partial^{2}}{\partial\lambda^{2}_{i}}+NV(\lambda_{i})\right\}\\
&+\frac{1}{4N}\sum_{i\neq
j}\frac{\hat{\tau}^{\mathcal{R}}_{ij}\hat{\tau}^{\mathcal{R}}_{ji}}{(\lambda_{i}-\lambda_{j})^{2}},\end{split}\end{equation}
where $P_{\mathcal{R}}$ is a projector for zero weight vectors in
the space of $\mathcal{R}$'th representation. A review of
$\mathcal{R}$'th representation can be found in Ref.\cite{Boulatov}.

\subsection{Non-singlet sector and T-duality}
\subsubsection{non-singlet terms from $U(N)$ generators}
As the previous case, the effective Lagrangian, $L_{eff}$, for $H$
is given by the Legendre transformation, and by the gaussian
integration for the $[d\Pi_{\lambda}]$ in following partition
function :
%eq-38
\begin{equation}\begin{split}&Z_{N}(b_{n})\\
&=\int[d\Pi_{\lambda}][d\lambda][d\chi_{ij}][d\bar{\chi}_{ij}]\\
&\times\exp\left[i\int dt\left(\sum_{i}^{N}\Pi_{\lambda_{i}}\dot{\lambda}_{i}+\sum_{ij}^{N}(\Pi_{\chi_{ij}}\dot{\chi}_{ji}+\Pi_{\bar{\chi}_{ij}}\dot{\bar{\chi}}_{ji})-H\right)\right]\\
&=\int[d\lambda][d\chi_{1}][d\chi_{2}]\exp\left(i\int dt
L_{eff}\right).\end{split}\end{equation} However, above equation is
not a correct form because the Hamiltonian does not contain the
non-singlet terms from the fermionic matrices involved in the
Lagrangian.

Earlier works on the adjoint state(non-singlet state) of the
supersymmetric matrix model can be found in Refs.\cite{Halpern01,
Halpern02}.

Now, we will consider the degrees of freedoms in the non-singlet
terms and the angular terms. Let us decompose $A(t)$ into generators
belonging to $U(N)$ as follows:
%eq-39
\begin{equation}A=\dot{U}U^{\dag}=\sum_{i=1}^{N-1}\dot{\alpha}C_{i}+\frac{i}{\sqrt{2}}\sum_{i\neq j}\bigl(\dot{\beta}_{ij}T_{ij}+\dot{\tilde{\beta}}_{ij}\widetilde{T}_{ij}\bigr),\end{equation}
where
%eq-40
\begin{equation}(T_{ij})_{kl}=\delta_{ik}\delta_{jl}+\delta_{il}\delta_{jk},\quad (\widetilde{T}_{ij})_{kl}=-i(\delta_{ik}\delta_{jl}-\delta_{il}\delta_{jk})\end{equation}
and $C_{i}$ is the Cartan subalgebra. If we insert Eq.(39) to
Eq.(25), we get
%eq-41
\begin{equation}L=\sum_{i=1}^{N}\left(\frac{1}{2}\dot{\lambda}_{i}^{2}-V\right)+\frac{1}{2}\sum_{i\neq j}(\lambda_{i}-\lambda_{j})^{2}\bigl(\dot{\beta}^{2}_{ij}+\dot{\tilde{\beta}}^{2}_{ij}\bigr).\end{equation}
The canonical relations,
%eq-42
\begin{equation}\Pi_{\lambda_{i}}=\frac{\partial L}{\partial\dot{\lambda}_{i}}\,,\qquad\Pi_{ij}=\frac{\partial L}{\partial\dot{\beta}_{ji}}\,,\qquad\widetilde{\Pi}_{ij}=\frac{\partial L}{\partial\dot{\tilde{\beta}}_{ji}}\end{equation}
and the constraint,
%eq-43
\begin{equation}\Pi_{\alpha_{i}}=\frac{\partial L}{\partial\dot{\alpha}_{i}}=0\end{equation}
with the Legendre transformation,
%eq-44
\begin{equation}H(q_{Iij}, \Pi_{Iij},
t)=\sum_{I}\Pi_{Iij}\dot{q}_{Iji}-L(q_{Iij}, \dot{q}_{Iij},
t)\end{equation} give us following Hamiltonian,
%eq-45
\begin{equation}H=\sum_{i}^{N}\left(\frac{1}{2}\Pi_{\lambda_{i}}^{2}+V(\lambda)\right)+\frac{1}{2}\sum_{i\neq j}\frac{\Pi^{2}_{ij}+\widetilde{\Pi}^{2}_{ij}}{(\lambda_{i}-\lambda_{j})^{2}}.\end{equation}
The second part in Eq.(45) is the interaction terms from the
non-singlet sector. In the coordinate representation the momentum is
realized as the following operator
%eq-46
\begin{equation}\Pi_{\lambda_{i}}=-\frac{i\hbar}{\Delta(\lambda)}\frac{\partial}{\partial\lambda_{i}}\Delta(\lambda).\end{equation}
Therefore, Eq.(45) becomes
%eq-47
\begin{equation}H=\sum_{i}^{N}\left[-\frac{\hbar^{2}}{2\Delta(\lambda)}\frac{\partial^{2}}{\partial\lambda_{i}^{2}}\Delta(\lambda)+V(\lambda)\right]+\frac{1}{2}\sum_{i\neq j}\frac{\Pi^{2}_{ij}+\widetilde{\Pi}^{2}_{ij}}{(\lambda_{i}-\lambda_{j})^{2}}.\end{equation}
Using the Hamiltonian derived in the previous paragraph, the
partition function becomes
%eq-48
\begin{equation}Z_{N}=\text{Tr}\,e^{-\frac{t}{\hbar}H},\end{equation}
where $t$ is the time interval we are interested in.

\subsubsection{T-duality and neglecting of non-singlet sector}
In this section, we will consider the relation of the non-singlet
terms and the T-duality. Details can be found in Refs.\cite{Gross0,
Gross1, Gross2}. The 2-dimensional target space has a time direction
and a spatial one. This spatial dimension is from the Liouville
modes so the spatial direction can not be compactified. Thus we
consider the time direction of the target space for the
compactification.

Firstly, let us consider a non-compactified target space. In this
case, we have a infinite real line for the time direction. The free
energy of this target space is sufficient to show that we need only
the ground state energy, $E_{0}$, as follows :
%eq-49
\begin{equation}F=\lim_{t\rightarrow\infty}\frac{\text{log}Z_{N}}{t}=-\frac{E_{0}}{\hbar}.\end{equation}
Above minimum value of $F$ is caused by fact that the angular modes
from the non-singlet terms can be decoupled. This decoupling arises
from fact that the non-singlet terms are positive definite
operators. So, we can consider following Hamiltonian of the singlet
sector which is independent of the angular degrees of freedom which
is related to the non-singlet terms,
%eq-50
\begin{equation}H_{singlet}=\sum_{i}^{N}\left(\frac{1}{2}\Pi_{\lambda_{i}}^{2}+V(\lambda)\right).\end{equation}

Now, let us consider a compactified time direction of the target
space as follows :
%eq-51
\begin{equation}t\sim t+\beta,\qquad\beta=2\pi R,\end{equation}
where {\it R} is the radius of the compactified target space. In
this case, we can not decouple the non-singlet terms because the
compactification gives rise to new winding modes which are
represented by the non-singlet terms mathematically. These winding
modes are related to the vortices and the anti-vortices. The
vortices and the anti-vortices bring about the vortex-anti-vortex
condensation which is related to the Kosterlitz-Thouless phase
transition at critical radius, $R_{c}$. However, this phase
transition violates the T-duality of the target space.

When the radius of the target space is big enough to be $R>R_{c}$,
we can suppress the phase transition. This means that we can
decouple and truncate the non-singlet terms under restriction of
$R>R_{c}$. The decoupling is caused by following relation,
%eq-52
\begin{equation}E_{non-singlet}-E_{singlet}\sim\frac{\beta}{2\pi}|\ln\delta|,\end{equation}
where $\delta\rightarrow0$ as $R\rightarrow\infty$. Thus we can
truncate the non-singlet sector because the energy gap between
$E_{non-singlet}$ and $E_{singlet}$ diverges to the infinity.

When the radius is small such that $R<R_{c}$, one can resort to the
lattice gauge theory. In this case, we can also suppress the
vortices and anti-vortices. However, the lattice gauge theory
transforms the circle of radius $R$ into the 1-dimensional lattice.
This modification of the target space is a restrictive condition.

\section{A new method for non-singlet sector in supersymmetric matrix quantum mechanics}
Till now, we reviewed the supersymmetric matrix quantum mechanics
and the bosonic matrix with time periodicity. We also reviewed the
compactification of the target space and reviewed the non-singlet
sector corresponding to the vortices which violate the T-duality.
The compactification gives us the time periodicity.

The non-singlet terms from the angular operators and from the time
periodicity violate the T-duality of the target space. However, when
we want to maintain the T-duality, we must eliminate, or ignore at
least, the non-singlet terms. We have seen, however, that we can
only ignore the non-singlet terms under some restrictions. The first
method is to use truncation with large difference of energy gab
between singlet states and non-singlet states. The second method is
to use the lattice gauge theory. However, this lattice gauge theory
transforms the circle of the target space into the 1-dimensional
lattice.

We will introduce a new method for the non-singlet terms. Our new
suggestion is not to ignore the non-singlet terms but to eliminate
of the non-singlet terms.

Firstly, let us look into the trivial singlet case of the fermionic
sector. We consider the diagonalized fermionic matrices by the
unitary transformation such as
%eq-53
\begin{equation}\Psi_{ij}(t)=\sum_{k}U^{\dag}_{ik}(t)\,\chi_{k}(t)\,U_{kj}(t).\end{equation}
The kinetic energy parts of Eq.(53) become as follows :
%eq-54
\begin{equation}\begin{split}
\sum_{ij}\Psi_{ij}\dot{\Psi}_{ji}&=\sum_{ij}\sum_{kl}U^{\dag}_{ik}\chi_{k}U_{kj}\cdot\frac{\partial}{\partial
t}\left(U^{\dag}_{jl}\chi_{l}U_{li}\right)\\
&=\sum_{k}\chi_{k}\dot{\chi}_{k}+2\sum_{ik}\chi^{2}_{k}\dot{U}_{ki}U^{\dag}_{ik}\\
&=\sum_{k}\chi_{k}\dot{\chi}_{k}+2\sum_{k}\chi^{2}_{k}A_{k}\\
&=\sum_{k}\chi_{k}\dot{\chi}_{k}\end{split}\end{equation} where
$A_{k}(t)=\bigl(\dot{U}(t)U^{\dag}(t)\bigr)_{kk}$ and
$\chi^{2}_{k}=\chi_{k}\chi_{k}=0$ because of the Pauli exclusion
principle. Thus, in this case we have no non-trivial interaction
terms for the fermionic parts. Next, we will investigate the
non-singlet terms of the fermionic parts in following subsection.

\subsection{Fermions in non-singlet sector}
Now, we consider the off-diagonal elements of the fermions in
non-singlet sector as follows :
%eq-55
\begin{equation}\Psi_{ij}(t)=\sum_{kl}U^{\dag}_{ik}(t)\chi_{kl}(t)U_{lj}(t).\end{equation}
The kinetic energy parts of Eq.(55) become such as
%eq-56
\begin{equation}\begin{split}
\sum_{ij}\Psi_{ij}\dot{\Psi}_{ji}&=\sum_{ij}\sum_{kl,mn}U^{\dag}_{ik}\chi_{kl}U_{lj}\cdot\frac{\partial}{\partial
t}\left(U^{\dag}_{jm}\chi_{mn}U_{ni}\right)\\
&=\sum_{kl}\chi_{kl}\dot{\chi}_{lk}+2\sum_{\substack{klni\\(k\neq
l\neq n)}}\chi_{kl}\chi_{ln}\dot{U}_{ni}U^{\dag}_{ik}\\
&=\sum_{kl}\chi_{kl}\dot{\chi}_{lk}+2\sum_{k\neq l\neq
n}\chi_{kl}\chi_{ln}A_{nk}.\end{split}\end{equation} Thus, we have
the non-trivial interaction terms from $\chi_{kl}\chi_{ln}A_{nk}$
for $k\neq l\neq n$. Note that the non-trivial terms have $A_{ij}$.
We will use this fact in the gaussian integral for $A_{ij}$ later.
%The superpotential terms don't contain $A_{ij}$.
Let us write the terms which have $A_{ij}$, as follows :
%eq-57
\begin{equation}\frac{1}{2}\text{Tr}\dot{M}^{2}=\frac{1}{2}\sum^{N}_{i}\dot{\lambda}^{2}_{i}+\frac{1}{2}\sum_{i\neq
j}(\lambda_{i}-\lambda_{j})^{2}|A_{ij}|^{2},\end{equation} and
%eq-58
\begin{equation}\begin{split}-&\frac{i}{2}\sum^{2}_{I}\sum_{ij}\Psi_{Iij}\dot{\Psi}_{Iji}\\
&=-\frac{i}{2}\sum^{2}_{I}\sum_{ij}\chi_{Iij}\dot{\chi}_{Iji}-i\sum^{2}_{I}\sum_{i\neq
k\neq j}\chi_{Iik}\chi_{Ikj}A_{ji}.\end{split}\end{equation} If we
use Eq.(57) and Eq.(58), the previous Lagrangian of Eq.(6) becomes
%eq-59
\begin{equation}\begin{split}L&=\frac{1}{2}\left\{\sum_{i}^{N}\dot{\lambda}^{2}_{i}+\sum_{i\neq j}(\lambda_{i}-\lambda_{j})^{2}|A_{ij}|^{2}\right.\\
&\left.-\sum_{ij}\frac{\partial W(M)}{\partial M_{ij}}\frac{\partial
W(M)}{\partial M_{ji}}\right\}-\frac{i}{2}\sum_{ij}(\chi_{1ij}\dot{\chi}_{1ji}+\chi_{2ij}\dot{\chi}_{2ji})\\
&-i\sum_{ijkl}\Psi_{1ij}\frac{\partial^{2}W(M)}{\partial
M_{ij}\partial M_{kl}}\Psi_{2kl}\\
&-i\sum_{i\neq k\neq
j}(\chi_{1ik}\chi_{1kj}+\chi_{2ik}\chi_{2kj})A_{ji}\end{split}\end{equation}
where those potential terms remain off the unitary transformation.

\subsection{Non-singlet terms from time periodicity}
In the section II.2 considered the periodic time condition on the
bosonic matrix case. Here we extend this time periodicity to the
supersymmetric matrix model. However, in the supersymmetric case,
using the gaussian integral for the $A_{ij}$ with the periodic time
condition, we have the same result of bosonic model. The same
process in the section II.2 gives us following Lagrangian, instead
of Eq.(59),
%eq-60
\begin{equation}\begin{split}L_{eff}&=\frac{1}{2}\sum_{i}^{N}\dot{\lambda}^{2}_{i}-\frac{i}{2}\sum_{ij}^{N}(\chi_{1ij}\dot{\chi}_{1ji}+\chi_{2ij}\dot{\chi}_{2ji})\\
&-\frac{1}{2}\sum_{i}^{N}\sum^{K}_{mn=1}mnb_{m}b_{n}\lambda^{m+n-2}_{i}\\
&-i\sum^{N}_{ij}\sum^{K}_{n=2}n(n-1)b_{n}\chi_{1ij}\lambda_{j}^{n-2}\chi_{2ji}\\
&-i\sum^{K}_{m=2}\sum^{m}_{n=2}\sum_{i\neq
j}mb_{m}\chi_{1ij}\lambda_{i}^{m-n}\lambda_{j}^{n-2}\chi_{2ji}\\
&+\frac{1}{2}\sum_{i\neq k\neq
j}\frac{(\sum_{\mathcal{R}}\hat{\tau}^{\mathcal{R}}_{ij}+\chi_{1ik}\chi_{1kj}+\chi_{2ik}\chi_{2kj})^{2}}{(\lambda_{i}-\lambda_{j})^{2}},\end{split}\end{equation}
where the potential part is rewritten in forms of the matrix
elements after unitary transformation and differentiation.

Consequently, we conclude that the non-singlet terms from the time
periodicity and the non-singlet terms for fermions give us same
effects. If we use the complex formation for the fermions like
Eq.(10),
%eq-61
\begin{equation}\begin{split}\chi_{ij}=&\frac{1}{\sqrt{2}}(\chi_{1ij}+i\chi_{2ij})
\\
\bar{\chi}_{ij}=&\frac{1}{\sqrt{2}}(\chi_{1ij}-i\chi_{2ij}),\end{split}\end{equation}
then we have
%eq-62
\begin{equation}\begin{split}L_{eff}=&\frac{1}{2}\sum_{i}^{N}\dot{\lambda}^{2}_{i}-\frac{i}{2}\sum_{ij}^{N}(\chi_{ij}\dot{\bar{\chi}}_{ji}+\bar{\chi}_{ij}\dot{\chi}_{ji})\\
&-\frac{1}{2}\sum_{i}^{N}\sum^{K}_{mn=1}mnb_{m}b_{n}\lambda^{m+n-2}_{i}\\
&-\frac{1}{2}\sum^{N}_{ij}\sum^{K}_{n=2}n(n-1)b_{n}\lambda_{j}^{n-2}(\bar{\chi}_{ij}\chi_{ji}-\chi_{ij}\bar{\chi}_{ji})\\
&-\frac{1}{2}\sum^{K}_{m=2}\sum^{m}_{n=2}\sum_{i\neq
j}mb_{m}\lambda_{i}^{m-n}\lambda_{j}^{n-2}(\bar{\chi}_{ij}\chi_{ji}-\chi_{ij}\bar{\chi}_{ji})\\
&+\frac{1}{2}\sum_{i\neq k\neq
j}\frac{(\sum_{\mathcal{R}}\hat{\tau}^{\mathcal{
R}}_{ij}+\chi_{ik}\bar{\chi}_{kj}+\bar{\chi}_{ik}\chi_{kj})^{2}}{(\lambda_{i}-\lambda_{j})^{2}}.\end{split}\end{equation}
Now, let us investigate a Hamiltonian for above the Lagrangian.

\subsection{Hamiltonian with fermionic non-singlet terms}
The two ways which ignore the non-singlet terms, have the restricted
conditions respectively\cite{Gross0, Gross1, Gross2}. How do we get
a more constructive and complete method for a eliminating the
non-singlet terms which are related to the vortices? A possible
answer is a using of the off-diagonal elements of the fermionic
matrices in the supersymmetric matrix model.

Now, we have three types of the non-singlet terms. The first type is
$\Pi_{ij}$, the second is $\hat{\tau}_{ij}$ and the third is
$\chi_{ik}\bar{\chi}_{kj}$ as follows :
\begin{enumerate}
  \item The $\Pi_{ij}$ is from the angular variable such as the $A_{ij}=(\dot{U}U^{\dag})_{ij}$.
  \item The $\hat{\tau}_{ij}$ is from the time periodicity.
  \item The $\chi_{ik}\bar{\chi}_{kj}$ is from the non-diagonal
  elements of the fermions.
\end{enumerate}
Let us consider following Lagrangian for a final Hamiltonian.
%eq-63
\begin{equation}\begin{split}L=&\frac{1}{2}\sum_{i}^{N}\dot{\lambda}^{2}_{i}+\frac{1}{2}\sum_{i\neq j}(\lambda_{i}-\lambda_{j})^{2}|A_{ij}|^{2}-V(\lambda, \chi, \bar{\chi})\\
&-\frac{i}{2}\sum_{ij}^{N}(\chi_{ij}\dot{\bar{\chi}}_{ji}+\bar{\chi}_{ij}\dot{\chi}_{ji})\\
&-i\sum_{i\neq k\neq
j}(\sum_{\mathcal{R}}\hat{\tau}^{\mathcal{R}}_{ij}+\chi_{ik}\bar{\chi}_{kj}+\bar{\chi}_{ik}\chi_{kj})A_{ji},\end{split}\end{equation}
where the potential, $V(\lambda, \chi, \bar{\chi})$, is
%eq-64
\begin{equation}\begin{split}V(\lambda, \chi,& \bar{\chi})=\frac{1}{2}\sum_{i}^{N}\sum^{K}_{mn=1}mnb_{m}b_{n}\lambda^{m+n-2}_{i}\\
&+\frac{1}{2}\sum^{N}_{ij}\sum^{K}_{n=2}n(n-1)b_{n}\lambda_{j}^{n-2}(\bar{\chi}_{ij}\chi_{ji}-\chi_{ij}\bar{\chi}_{ji})\\
&+\frac{1}{2}\sum^{K}_{m=2}\sum^{m}_{n=2}\sum_{i\neq
j}mb_{m}\lambda_{i}^{m-n}\lambda_{j}^{n-2}(\bar{\chi}_{ij}\chi_{ji}-\chi_{ij}\bar{\chi}_{ji}).\end{split}\end{equation}
As previous case, let us decompose $A(t)$ into generators belonging
to $U(N)$ as follows :
%eq-65
\begin{equation}A=\dot{U}U^{\dag}=\sum_{i=1}^{N-1}\dot{\alpha}C_{i}+\frac{i}{\sqrt{2}}\sum_{i\neq j}\bigl(\dot{\beta}_{ij}T_{ij}+\dot{\tilde{\beta}}_{ij}\widetilde{T}_{ij}\bigr),\end{equation}
where
%eq-66
\begin{equation}(T_{ij})_{kl}=\delta_{ik}\delta_{jl}+\delta_{il}\delta_{jk},\quad (\widetilde{T}_{ij})_{kl}=-i(\delta_{ik}\delta_{jl}-\delta_{il}\delta_{jk})\end{equation}
and $C_{i}$ is Cartan subalgebra. In Eq.(63), the summation
conditions, $\sum_{i\neq j}$ and $\sum_{i\neq k\neq j}$, make
following constraint,
%eq-67
\begin{equation}\Pi_{\alpha_{i}}=\frac{\partial L}{\partial\dot{\alpha}_{i}}=0,\end{equation}
thus we have
%eq-68
\begin{equation}A_{ij}=(\dot{U}U^{\dag})_{ij}=\frac{i}{\sqrt{2}}\sum_{i\neq j}\bigl(\dot{\beta}_{ij}T_{ij}+\dot{\tilde{\beta}}_{ij}\widetilde{T}_{ij}\bigr).\end{equation}
Now, the description of Eq.(68) is comparatively complicated. Since
the term $A_{ij}$ which definite form is
$A_{ij}(t)=(\dot{U}(t)U^{\dag}(t))_{ij}$, has time derivative part,
we can redefine the $A_{ij}$ as follows :
%eq-69
\begin{equation}A_{ij}\equiv \dot{\gamma}_{ij},\end{equation} then we have following canonical relations instead of Eq.(42),
%eq-70
\begin{equation}\Pi_{\lambda_{i}}=\frac{\partial L}{\partial\dot{\lambda}_{i}},\qquad\Pi_{\gamma_{ij}}=\frac{\partial
L}{\partial\dot{\gamma}_{ji}}.\end{equation} Comparing Eq.(42) to
Eq.(70) we have following relation,
%eq-71
\begin{equation}\Pi^{2}_{\gamma_{ij}}=\Pi^{2}_{ij}+\widetilde{\Pi}^{2}_{ij}.\end{equation}
From now on, we relabel $\Pi_{\gamma_{ij}}$ as follows :
%eq-72
\begin{equation}\Pi_{\gamma_{ij}}=\widehat{\Pi}_{ij}.\end{equation}
Now, we have a following Lagrangian,
%eq-73
\begin{equation}\begin{split}L=&\frac{1}{2}\sum_{i}^{N}\dot{\lambda}^{2}_{i}+\frac{1}{2}\sum_{i\neq j}(\lambda_{i}-\lambda_{j})^{2}|\dot{\gamma}_{ij}|^{2}\\
&-V(\lambda, \chi, \bar{\chi})\\
&-\frac{i}{2}\sum_{ij}^{N}(\chi_{ij}\dot{\bar{\chi}}_{ji}+\bar{\chi}_{ij}\dot{\chi}_{ji})\\
&-i\sum_{i\neq k\neq
j}(\sum_{\mathcal{R}}\hat{\tau}^{\mathcal{R}}_{ij}+\chi_{ik}\bar{\chi}_{kj}+\bar{\chi}_{ik}\chi_{kj})\dot{\gamma}_{ji}\end{split}\end{equation}
and the canonical relations as follows :
%eq-74
\begin{equation}\begin{split}&\Pi_{\lambda_{i}}=\frac{\partial
L}{\partial\dot{\lambda}_{i}}=\dot{\lambda}_{i}\\
&\Pi_{\chi_{ij}}=\frac{\partial
L}{\partial\dot{\chi}_{ji}}=-\frac{i}{2}\bar{\chi}_{ij}\\
&\Pi_{\bar{\chi}_{ij}}=\frac{\partial
L}{\partial\dot{\bar{\chi}}_{ji}}=-\frac{i}{2}\chi_{ij}\\
&\widehat{\Pi}_{ij}=\frac{\partial
L}{\partial\dot{\gamma}_{ji}}=(\lambda_{i}-\lambda_{j})^{2}\dot{\gamma}_{ij}\\
&\qquad\qquad\qquad-i(\sum_{\mathcal{R}}\hat{\tau}^{\mathcal{R}}_{ij}+\chi_{ik}\bar{\chi}_{kj}+\bar{\chi}_{ik}\chi_{kj}).\end{split}\end{equation}
Thus, we have following description for $\dot{\gamma}$,
%eq-75
\begin{equation}\dot{\gamma}_{ij}=\frac{\widehat{\Pi}_{ij}+i(\sum_{\mathcal{R}}\hat{\tau}^{\mathcal{R}}_{ij}+\chi_{ik}\bar{\chi}_{kj}+\bar{\chi}_{ik}\chi_{kj})}{(\lambda_{i}-\lambda_{j})^{2}}.\end{equation}
Inserting Eq.(75) into Eq.(73) then we have following Lagrangian
%eq-76
\begin{equation}\begin{split}L=&\sum_{i}^{N}\frac{1}{2}\dot{\lambda}_{i}^{2}-V(\lambda, \chi, \bar{\chi})-\frac{i}{2}\sum_{ij}^{N}(\chi_{ij}\dot{\bar{\chi}}_{ji}+\bar{\chi}_{ij}\dot{\chi}_{ji})\\
&+\frac{1}{2}\sum_{i\neq
j}\frac{\widehat{\Pi}_{ij}\widehat{\Pi}_{ji}}{(\lambda_{i}-\lambda_{j})^{2}}\\
&+\frac{1}{2}\sum_{i\neq
k\neq j}\frac{(\sum_{\mathcal{R}}\hat{\tau}^{\mathcal{R}}_{ij}+\chi_{ik}\bar{\chi}_{kj}+\bar{\chi}_{ik}\chi_{kj})^{2}}{(\lambda_{i}-\lambda_{j})^{2}}\\
&+\frac{i}{2}\sum_{i\neq k\neq
j}\frac{\widehat{\Pi}_{ij}(\sum_{\mathcal{R}}\hat{\tau}^{\mathcal{R}}_{ji}+\chi_{jk}\bar{\chi}_{ki}+\bar{\chi}_{jk}\chi_{ki})}{(\lambda_{i}-\lambda_{j})^{2}}\\
&-\frac{i}{2}\sum_{i\neq k\neq
j}\frac{(\sum_{\mathcal{R}}\hat{\tau}^{\mathcal{R}}_{ij}+\chi_{ik}\bar{\chi}_{kj}+\bar{\chi}_{ik}\chi_{kj})\widehat{\Pi}_{ji}}{(\lambda_{i}-\lambda_{j})^{2}}.\end{split}\end{equation}
With above Lagrangian, the canonical relations of Eq.(74), and the
Legendre transformation,
%eq-77
\begin{equation}H(q_{Iij}, \Pi_{Iij},
t)=\sum_{I=1}^{4}\Pi_{Iij}\dot{q}_{Iji}-L(q_{Iij}, \dot{q}_{Iij},
t),\end{equation} where
%eq-78
\begin{equation}\sum_{I=1}^{4}\Pi_{Iij}\dot{q}_{Iji}=\Pi_{\lambda_{i}}\dot{\lambda_{i}}+\Pi_{\chi_{ij}}\dot{\chi}_{ji}+\Pi_{\bar{\chi}_{ij}}\dot{\bar{\chi}}_{ji}+\widehat{\Pi}_{ij}\dot{\gamma}_{ji},\end{equation}
we have a extended Hamiltonian in the supersymmetric and periodic
time (compactified target space) case,
%eq-79
\begin{equation}\begin{split}H&=\sum_{ij}\left(\Pi_{\lambda_{i}}\dot{\lambda_{i}}+\Pi_{\chi_{ij}}\dot{\chi}_{ji}+\Pi_{\bar{\chi}_{ij}}\dot{\bar{\chi}}_{ji}+\widehat{\Pi}_{ij}\dot{\gamma}_{ji}\right)-L\\
&=\sum_{i}\Pi_{\lambda_{i}}^{2}-\frac{i}{2}\sum_{ij}(\chi_{ij}\dot{\bar{\chi}}_{ji}+\bar{\chi}_{ij}\dot{\chi}_{ji})-L\\
&+\sum_{i\neq k\neq
j}\frac{\widehat{\Pi}_{ij}\widehat{\Pi}_{ji}+i\widehat{\Pi}_{ij}(\sum_{\mathcal{R}}\hat{\tau}^{\mathcal{R}}_{ji}+\chi_{jk}\bar{\chi}_{ki}+\bar{\chi}_{jk}\chi_{ki})}{(\lambda_{i}-\lambda_{j})^{2}}.\end{split}\end{equation}
Inserting the Lagrangian of Eq.(76), into above Eq.(79), we have
following result,
%eq-80
\begin{equation}\begin{split}H&=\frac{1}{2}\sum_{i}\Pi_{\dot{\lambda}_{i}}^{2}+V(\lambda, \chi,
\bar{\chi})+\frac{1}{2}\sum_{i\neq
j}\frac{\widehat{\Pi}_{ij}\widehat{\Pi}_{ji}}{(\lambda_{i}-\lambda_{j})^{2}}\\
&\qquad-\frac{1}{2}\sum_{i\neq
k\neq j}\frac{(\sum_{\mathcal{R}}\hat{\tau}^{\mathcal{R}}_{ij}+\chi_{ik}\bar{\chi}_{kj}+\bar{\chi}_{ik}\chi_{kj})^{2}}{(\lambda_{i}-\lambda_{j})^{2}}\\
&\qquad+\frac{i}{2}\sum_{i\neq k\neq
j}\frac{\widehat{\Pi}_{ij}(\sum_{\mathcal{R}}\hat{\tau}^{\mathcal{R}}_{ji}+\chi_{jk}\bar{\chi}_{ki}+\bar{\chi}_{jk}\chi_{ki})}{(\lambda_{i}-\lambda_{j})^{2}}\\
&\qquad+\frac{i}{2}\sum_{i\neq k\neq
j}\frac{(\sum_{\mathcal{R}}\hat{\tau}^{\mathcal{R}}_{ij}+\chi_{ik}\bar{\chi}_{kj}+\bar{\chi}_{ik}\chi_{kj})\widehat{\Pi}_{ji}}{(\lambda_{i}-\lambda_{j})^{2}}.\end{split}\end{equation}
At last, with some calculation and rearrangement of the terms in
above equation, we arrive at this final Hamiltonian form,
%eq-81
\begin{equation}\begin{split}H&=\frac{1}{2}\sum_{i}^{N}\Pi_{\lambda_{i}}^{2}+V(\lambda, \chi,
\bar{\chi})\\
&+\frac{1}{2}\sum_{i\neq k\neq
j}\frac{[\widehat{\Pi}_{ij}+i(\sum_{\mathcal{R}}\hat{\tau}^{\mathcal{R}}_{ij}+\chi_{ik}\bar{\chi}_{kj}+\bar{\chi}_{ik}\chi_{kj})]^{2}}{(\lambda_{i}-\lambda_{j})^{2}}.\end{split}\end{equation}
Notice that the last four terms of Eq.(80) have been collected into
a perfect square and the Hamiltonian is simplified. It is rather
remarkable that the fermionic non-singlet terms and the non-singlet
terms from the time periodicity and the angular variable conspire to
give a simple form.

Here, the terms of $\widehat{\Pi}_{ij}$ and $\hat{\tau}_{ij}$ are
not controlled by us since they are given from the structure of the
matrix model and some mathematical conditions. For example, the
$\widehat{\Pi}_{ij}$ is from the angular variable of the $A_{ij}$
and the unitary transformation. Similarly, the $\hat{\tau}_{ij}$ is
from the time periodicity. We can't change these restricted
conditions arbitrarily. However, we can control and vary the terms
of $\chi$ and $\bar{\chi}$ since they are from the superfields of
Eq.(1), which are introduced by us.

By the way, the non-singlet terms which are made of the
$\widehat{\Pi}_{ij}$ and/or $\hat{\tau}_{ij}$, violate the T-duality
of the target space\cite{Gross0, Gross1, Gross2}. But, if we want to
maintain the T-duality then we must suppress the non-singlet terms.
Really, we would like to retain the T-duality because that the
T-duality is very good symmetry for us. So, the non-singlet terms of
the $\widehat{\Pi}_{ij}$ and/or $\hat{\tau}_{ij}$ are suppressed and
ignored by the two ways in the previous papers\cite{Gross0, Gross1,
Gross2}. However, as remarked above the review section, the two ways
have some restrictive conditions respectively.

Therefore, if we redefine the numerator of the non-singlet terms as
follows :
%eq-82
\begin{equation}\widehat{\Pi}_{ij}+i(\sum_{\mathcal{R}}\hat{\tau}^{\mathcal{R}}_{ij}+\chi_{ik}\bar{\chi}_{kj}+\bar{\chi}_{ik}\chi_{kj})\equiv \Omega_{ij},\end{equation}
then we have following description of the Hamiltonian,
%eq-83
\begin{equation}H=\frac{1}{2}\sum_{i}^{N}\Pi_{\lambda_{i}}^{2}+V(\lambda, \chi,
\bar{\chi})+\frac{1}{2}\sum_{i\neq
j}\frac{\Omega_{ij}\Omega_{ji}}{(\lambda_{i}-\lambda_{j})^{2}}.\end{equation}
Now, for the T-duality, all we want is not ignoring but complete
elimination of the non-singlet terms which violate the T-duality.
So, if we control and vary the terms of
$\chi_{ik}\bar{\chi}_{kj}+\bar{\chi}_{ik}\chi_{kj}$ for
$\Omega_{ij}=0$ in Eq.(82) then we can have following Hamiltonian,
%eq-84
\begin{equation}H=\frac{1}{2}\sum_{i}^{N}\Pi_{\lambda_{i}}^{2}+V(\lambda, \chi,
\bar{\chi}),\end{equation} which retain the T-duality of the target
space. With a using of the $\chi_{ij}$ and $\bar{\chi}_{ij}$, this
elimination of the non-singlet terms is not under any restrictive
conditions and constraints.

Now, we can have the Hamiltonian without the non-singlet terms which
violate the T-duality even in the target space of a circle of
arbitrary radius $R$. This fact means that we can eliminate the
vortices and anti-vortices on the target space without any
constraint. Therefore we can always retain the T-duality on the
target space of a circle.

\section{Discussion}
The target space of the 1-dimensional(time dimension) matrix model
related to the non-critical 2-dimensional string theory, has three
structures such as the infinite real line, the infinite
1-dimensional lattice and the circle of radius {\it R}. The target
space can be represented by discretised random surfaces and have a
dual structure of the fat Feynman graphs. If we consider the
non-singlet sector in the matrix model then we can read the
non-singlet terms into vortex or anti-vortex terms. Also these
non-singlet terms correspond to the winding modes of the strings.

However, in the continuum limit and the double scaling limit on the
random surfaces, the Kosterlitz-Thouless phase transition through
the vortex-anti-vortex condensation, which violate the T-duality of
the target space. In the case such that the target space is infinite
line or infinite 1-dimensional lattice, we need not to consider the
non-singlet terms since we need only ground state energy. Since the
non-singlet terms are positive definite operators, corresponding
excitation state energy is always above the ground state one. This
excitation states occur where the target space is a circle of radius
{\it R}. In general, the compact target space with radius {\it R}
has T-duality. But the target space in our case is composed of
discretised random surfaces. Therefore we have vortex or anti-vortex
terms on the surfaces. The non-singlet terms corresponding to the
vortices and anti-vortices which violate the T-duality of the target
space. Up to now, we have two old ways\cite{Gross0, Klebanov,
Gross1, Gross2} to exclude the violation of the T-duality. Firstly,
in the continuum limit of discretised surfaces, we are able to
truncate the vortex terms since the energy gap between ground state
and excitation state diverges. This method corresponds to the
infinite limit of the radius {\it R} of the target space. Secondly,
using the lattice gauge theory, we are able to read the circle into
the 1-dimensional lattice. However these two ways are in restrictive
conditions respectively.

In this paper we showed new non-singlet terms from the non-diagonal
elements of the fermionic matrices in the Lagrangian. These new
non-singlet terms can completely eliminate the old non-singlet terms
which violate the T-duality so that we can retain the T-duality on
the target space which is composed of the discretised random
surfaces. What is more, we are also able to control the phase
transition effect with these new non-singlet terms which are
composed of $\chi$ and $\bar{\chi}$ instead of the elimination.

\begin{acknowledgments}
The work of S.Nam is supported by grant No. R01-2003-000-10391-0
from the Basic Research Program of the Korea Science and Engineering
Foundation and by the Center for Quantum Spacetime (CQUeST) of
Sogang University with grant number R11-2005-021(KOSEF).

The work of Y.Seo is supported by grant No. R01-2004-000-10520-0
from the Basic Research Program of the Korea Science and Engineering
Foundation.
\end{acknowledgments}


\begin{thebibliography}{99}
\bibitem{GG} P. Ginsparg and Gregory Moore, {\it Lectures on 2D Gravity and 2D String Theory}, Lectures given June 11-19, 1992 at TASI summer school, Boulder, CO
\bibitem{FGZ} P. Di Francesco, P. Ginsparg and J. Zinn-Justin, Phys. Rept. {\bf 254} 1-133 (1995)
\bibitem{Polychronakos} Alexios P. Polychronakos, {\it Generalized statistics in one dimension}, Les Houches 1998 Lectures, [hep-th/9902157].
\bibitem{Gross0} David J. Gross, {\it The c=1 matrix models},
(1990.12$\sim$1991.01) Jerusalem winter school for theoretical
physics, {\it Two dimensional quantum gravity and random surfaces},
Ed by D. J. Gross, T. Piran and S. Weinberg, World Scientific,
(1992) 143-173.
\bibitem{Klebanov} Igor R. Klebanov, {\it String Theory in Two Dimensions}, (1991) Lectures delivered at the ICTP Spring School on String Theory and Quantum Gravity, Trieste.
\bibitem{Das} Sumit R. Das, {\it The One Dimensional Matrix Model and String Theory}, (1992) Lectures delivered at the Spring School on Superstrings at ICTP, Trieste.
\bibitem{Martinec} Emil J. Martinec, {\it Matrix Models and 2D String Theory}, (2004) Lectures given at the summer school ``Applications of Random Matrices in Physics", les Houches; and the Summer School on Strings, Gravity and Cosmology, Univ. of British Columbia.
\bibitem{Ovrut} Ram Brustein, Michael Faux and Burt A. Ovrut, Nucl. Phys. {\bf B421} 293-342 (1994).
\bibitem{Boulatov} Dmitri Boulatov and Vladimir Kazakov, Int. J. Mod. Phys. {\bf A8} 809-852 (1993).
\bibitem{KKK} Vladimir Kazakov, Ivan Kostov and David Kutasov, Nucl. Phys. {\bf B622} 141-188 (2002).
\bibitem{Halpern01} M. B. Halpern and C. Schwartz, {\it Large N classical solution for the one matrix model}, Phys. Rev. {\bf D24} 2146 (1981).
\bibitem{Halpern02} M. Claudson and M. B. Halpern, {\it Supersymmetric ground state wave functions}, Nucl. Phys. {\bf B250} 689 (1985).
\bibitem{Tonder} Joao P. Rodrigues and Andre J. van Tonder, Int. J. Mod. Phys. {\bf A8} 2517-2550 (1993).
\bibitem{Jevicki} Jean Avan and Antal Jevicki, Nucl. Phys. {\bf B469} 287-301 (1996).
\bibitem{Gross1} David J. Gross and Igor Klebanov, Nucl. Phys. {\bf
B344} 475-498 (1990).
\bibitem{Gross2} David J. Gross and Igor Klebanov, Nucl. Phys. {\bf
B354} 459-474 (1991).
\bibitem{Alexandrov} Sergei Alexandrov, {\it Matrix Quantum Mechanics and Two-dimensional String Theory in Non-trivial
Backgrounds}, [hep-th/0311273].
\bibitem{Koetsier} Arnaud Koetsier, {\it Matrix Models of 2D String
Theory in Non-trivial Backgrounds}, [hep-th/0509024].
\end{thebibliography}
\end{document}